\documentstyle[aps,prl,epsfig]{revtex}

\def\mb{MgB$_2$ }

\def\kp{ k^{(+)}_{ij} }
\def\km{ k^{(-)}_{ij} }

\def\etal{{\em et al}}

\newcommand{\beq}{\begin{equation}}
\newcommand{\unbeq}{\end{equation}}
\newcommand{\eeq}{\end{equation}}
\newcommand{\ba}{\begin{eqnarray}}
\newcommand{\unba}{\end{eqnarray}}
\newcommand{\ea}{\end{eqnarray}}

\begin{document}
\twocolumn[\hsize\textwidth\columnwidth\hsize\csname
@twocolumnfalse\endcsname

\title{\bf Transition Spectra for a BCS Superconductor 
with Multiple Gaps: Model Calculations for \mb}
\author{Sergey V. Barabash and David Stroud}
\address{
Department of Physics,
The Ohio State University, Columbus, Ohio 43210}

\date{\today}

\maketitle

\begin{abstract}
We analyze the qualitative features in the transition spectra 
of a model superconductor with multiple energy gaps, using a 
simple extension of the Mattis-Bardeen expression for probes with case I 
and case II coherence factors.  At temperature $T = 0$, the far infrared 
absorption edge is, as expected, determined by the smallest gap.  
However, the large  thermal background
may mask this edge at finite temperatures and instead the
secondary absorption edges found at 
$\Delta_i+\Delta_j$ may become most prominent.
At finite $T$, if certain interband matrix elements are large,
there may also be absorption peaks at the gap difference
frequencies $|\Delta_i-\Delta_j|/\hbar$. 
We discuss the effect of sample quality on the measured spectra and
the possible relation of these predictions to the recent 
infrared absorption measurement on \mb.
\end{abstract}

\draft 
\vskip1.5pc]

Among the unusual features of superconductivity in
\mb is the widely discussed possibility that there may
exist several distinct energy gaps.  
The larger gap or gaps are thought to be associated with electrons in
the boron $p_{xy}$ bands, frequently referred to as ``2D-bands'' 
because of their very weak dependence on $k_z$, the component of
${\bf k}$ parallel to the $c$ axis.  These bands appear to be 
strongly coupled to a 
particular ($E_{2g}$) phonon mode, and are believed primarily 
responsible 
for the high superconducting $T_c$ ($\sim 40$K).
Interactions between these 2D electrons and those in
other bands (specifically, the ``3D'' bands formed by the 
B $p_z$ states) lead to superconducting gaps in the spectra of
these other electrons.  A multiple-gap structure can be described 
either
with a simple BCS framework\cite{Liu}, or by more detailed
{\em ab initio} calculations using
the Eliashberg theory\cite{mb111823}.
The result of such theories is that the average 2D gap, 
$\Delta_{2D}$, is about three times larger than 
$\Delta_{3D}$\cite{note:2D}.  
These predictions are supported experimentally by
heat capacity \cite{mb111823,mb07072,mb11262}, 
tunneling \cite{mb12144,mb12452,mb11115,mb0201200}, 
photoemission\cite{tsuda01},
and penetration depth\cite{mbLemberger} measurements.

In this work, we present a simple model
calculation of the transition spectra in a superconductor with
multiple gaps, using a natural extension 
of the Mattis-Bardeen (MB) formulae\cite{MattisBardeen}.
We find that this model is not only consistent with the 
observed infrared absorption edge in \mb, which occurs at anomalously
low frequencies relative to the single-gap
BCS prediction,
but also predicts that a characteristic 
additional structure in the absorption
may be observed under certain conditions.

In our calculations, we consider only the effects of multiple
gaps on coherence phenomena, and on the superconducting
density of states, and neglect 
nonlocal electrodynamic effects.  Such an approximation is known to 
become exact in either the dirty limit 
($\ell \ll \xi_0$, where $\ell$ is
the quasiparticle mean-free path and 
$\xi_0$ is the coherence length), or
the ``extreme anomalous'' limit 
($k\xi_0 \gg 1$, where $k$ is the wave vector of the perturbation).
In fact, neither of these limits may actually be applicable to 
samples of \mb with multiple gaps (in particular,
a single gap would normally be expected in the dirty 
limit\cite{MazinCleanDirty}).
Nevertheless, previous calculations suggest that
the MB formalism gives at least a reasonable qualitative
description of experiment in single-gap superconductors, 
even when used beyond its nominal validity limit.
The calculation is also very easy.  
Thus, it is reasonable to do such a calculation
as a first step in understanding the absorption spectra in
MgB$_2$.  Various further corrections should be included
in the strong-coupling limit\cite{referee1a}, although
they may not change the calculated absorption spectra
qualitatively\cite{referee1b}.

We adopt a BCS-like multiple gap model proposed by 
Liu \etal\cite{Liu}.  In this model, the 
self-consistent gap equation is written as
\begin{equation}
\Delta_i = \sum_j U_{ij} \Delta_j
	\int\limits_{-\omega_D}^{\omega_D} N_j(\xi)
	 \frac{\tanh(\beta E_{j}(\xi)/2)}{E_{j}(\xi)}d\xi,
\label{eq:BCS}
\end{equation}
where $\beta = 1/k_BT$, with $T$ the temperature,
$\Delta_i$ is one of the $n$ gaps, 
$U_{ij}$ is the corresponding $n \times n$
effective phonon-mediated electron-electron interaction matrix 
calculated in \cite{Liu},
$N_i(\xi)$ is the normal density of states for the $i$th band, and
$ E_{i}(\xi) = \sqrt{\Delta_i^2 + \xi^2}$,
$\xi = \epsilon - \mu$ (where $\epsilon$ is
the single-particle energy and $\mu$ is the Fermi energy).
Finally, $\omega_D$ is a cutoff energy, which is assumed to be
the same for all $n$ bands.
The solution of the $n$ equations (\ref{eq:BCS}) 
gives $\Delta_i(T)$. 
In a single-gap BCS superconductor $\omega_D$ is 
of order the Debye frequency.  In \mb the value
$\omega_D\approx 7.5$meV needed to produce
$T_c \approx 40$K is much smaller than the physically relevant 
logarithmically averaged phonon frequency
$\omega_{\ln} =56.2$meV\cite{Liu}); this discrepancy would
probably be removed by the inclusion of strong-coupling corrections
omitted from eq. (\ref{eq:BCS}).
In any case, the BCS model
predicts gaps in \mb whose ratio and temperature dependence
agree fairly well with detailed calculations employing the Eliashberg 
theory of superconductivity\cite{mb111823}
(cf.\ Fig. 1(a) below).
We therefore use the model of eq.\ (1) to calculate absorption 
spectra in \mb, in the hope that the results will apply
at least qualitatively.

To calculate the absorption coefficients, we use the 
canonical transformation approach as described, for example, 
in Tinkham\cite{Tinkham}.  If there are multiple gaps,
one may generalize the standard single-gap expression
for the coherence factors $(u \nu^{\prime} \pm \nu u^\prime )^2$ and 
$(u u^\prime  \mp \nu\nu^\prime )^2$ by replacing 
$\Delta^2$ in these expressions by $\Delta \Delta^\prime$ (in the notation of
\cite{Tinkham}), so that, e.\ g.,
\beq
(u u^\prime  \mp \nu\nu^\prime )^2 = \frac 1 2 \left( 
	1 + \frac{\xi\xi^\prime }{E E^\prime } 
	\mp \frac{\Delta \Delta^\prime }{E E^\prime }
	\right).
\label{eq:factor}
\eeq
Here
the upper and lower signs correspond to the so-called case I and 
case II coherence factors determined by the time-reversal 
symmetry of the matrix elements. 
If the $N_j(\xi)$'s near the
Fermi level are constant for each band, and equal to $N_j(0)$,
and if the electron-phonon interaction is linear 
(both in $\xi$ and in the ionic displacements $u$), 
so that the upper and lower limits of integration in 
eq.\ (\ref{eq:BCS}) are equal, then
the terms linear in $\xi$ and $\xi^\prime $ cancel when the integral
over the coherence factors is carried out.
Both of these assumptions may fail to some extent in \mb 
(in particular, the electron-phonon interaction 
in \mb is exceptionally nonlinear in $u$\cite{Liu,Yildirim}).
We will nevertheless assume that the cancellation is almost complete 
and neglect the terms linear in $\xi$ and $\xi^\prime$ in eq.\ 
(\ref{eq:factor})\cite{note:cancel}.
We also assume that the matrix elements $M_{ij}$ for a one-electron
transition between an electronic state in band 
$i$ and a state in band $j$ are the 
same for all states in given bands $i$ and $j$.  
The transition rate in the superconducting state 
induced by a perturbation of frequency $\omega$ is then found to be 
proportional to
\ba
\alpha_S = &&\sum\limits_{ij}|M_{ij}|^2N_i(0)N_j(0) I_{ij}, \\
I_{ij}\equiv &&\int 
	\frac{ | E(E+\hbar\omega)\mp\Delta_i\Delta_j | }
	{ \sqrt{E^2-\Delta_i^2} \sqrt{(E+\hbar\omega)^2-\Delta_j^2} } 
\label{eq:Iij} \\
&&\,\,\,\,\,\,\,\,\,\,\,\,\,\,\,\,\,\,\,\,\,\,\,\,\,\,
  \times\left[f(E)-f(E+\hbar\omega)\right]dE.
\nonumber
\ea
Here $f(E) = 1/[e^{E/k_BT} + 1]$ is the Fermi function, 
and the integration in (\ref{eq:Iij}) extends from 
$-\infty$ to $+\infty$, except for
the regions where 
the argument of either square root becomes negative.
The absorption coefficient in the normal state, $\alpha_N$,
is given by the same formula but with all $\Delta_i = 0$.

In the single-gap case, the ratio $\alpha_S/\alpha_N$ calculated
from eqs.\ (3) and (4) is usually referred to
as the Mattis-Bardeen formula.
The corresponding ratio for the multiple-gap case is readily
calculated using eqs. (3) and (4).  The 
normal state value is just
$\alpha_N = \sum_{ij}|M_{ij}|^2N_i(0)N_j(0)\hbar\omega$, and hence
\begin{equation}
\frac{\alpha_S}{\alpha_N} = \frac{\sum_{i=1}^n\sum_{j=1}^n|M_{ij}|^2N_i(0)N_j(0)I_{ij}}
{\sum_{i=1}^n\sum_{j=1}^n|M_{ij}|^2N_i(0)N_j(0)\hbar\omega},
\label{eq:ratio}
\end{equation}
with $I_{ij}$ given by (\ref{eq:Iij}).

The required integrals take very simple forms at $T = 0$.
In this case, $f(E) = 1$ for $E <0 $ and $f(E) = 0$
otherwise, and  eq.\ (\ref{eq:Iij}) becomes
\begin{equation}
I_{ij}^{T=0} = \int_{\Delta_j-\hbar\omega}^{-\Delta_i}
	\frac{ | E(E+\hbar\omega) \mp \Delta_i\Delta_j | \,\, dE }
	{ \sqrt{E^2-\Delta_i^2} \sqrt{(E+\hbar\omega)^2-\Delta_j^2} } \,\,.
\label{eq:T0}
\end{equation}
Introducing
$ \kp \equiv (\hbar\omega)^2 - ( \Delta_i + \Delta_j)^2 $ and
$ \km \equiv (\hbar\omega)^2 - ( \Delta_i - \Delta_j)^2 $,
we can write this integral in terms of complete elliptic integrals $E$ and $K$.
For the case I coherence factors, (\ref{eq:T0}) simply reduces to
\beq
I_{ij}^{I,T=0} = \theta(\kp)
	\sqrt{\km} E \left( \frac\kp\km \right),
\eeq
whereas for the case II factors
\ba
I_{ij}^{II,T=0} = \theta(\kp)
	&&\left[ \sqrt{\km} 
			E \left( \frac\kp\km \right) \right. \nonumber \\
		&&\left. - \frac{4\Delta_i\Delta_j}{\sqrt{\km}} 
			K\left( \frac\kp\km \right) 
	\right]
\ea
[the step function $\theta(x)$ appears because the upper integration limit
in (\ref{eq:T0}) should always be larger than the lower one].

In order to demonstrate the qualitative features of this model,
we have carried out the numerical integration
for the two-gap model of \mb \cite{Liu,note:model}.  Even using 
this model,
the matrix elements $M_{ij}$ still remain to be calculated.
Now in MgB$_2$, these two gaps are thought to come from two
disconnected Fermi surfaces.  Since these are disjoint, the
normal-state resistivity is likely to be determined primarily
by {\em intraband} scattering.  If so, the {\em
off-diagonal} matrix elements $M_{ij}$ would be very small.
Various scattering processes (e. g. impurity-scattering,
electron-phonon scattering involving a large-wave-vector
phonon) could, however, produce non-zero off-diagonal 
matrix elements in principle.  We have therefore  
carried out two model calculations,
based on eq.\ (\ref{eq:ratio}.  In the first, we have simply
made the crude assumption 
that all the factors $M_{ij}N_i(0)N_j(0)$ are equal, so that the 
Eq.\ (\ref{eq:ratio}) becomes simply
\begin{equation}
\frac{\alpha_S}{\alpha_N} = \frac{\sum_{ij}I_{ij}}{n^2\hbar\omega},
\label{eq:ratioSimple}
\end{equation}
where $n$ is the number of bands.  In the second, we have
taken $M_{ij} = 0$ for $i \neq j$, as suggested by the above
argument, but still assuming all diagonal elements
$M_{ii}N_i(0)^2$ to be equal, so that
\begin{equation}
\frac{\alpha_S}{\alpha_N} = \frac{\sum_iI_{ii}}{n\hbar\omega}.
\label{eq:ratioDiag}
\end{equation}

In Figs.\ 1(b-e), we show the transition rates as
calculated from these expressions, for processes with 
case I (dashed lines) and case II (solid lines) coherence factors,
and making either of the assumptions (\ref{eq:ratioSimple}) or 
(\ref{eq:ratioDiag}) at $T = 0.05T_c$ and $T = 0.5T_c$,
as indicated.  The case II coherence factors correspond to 
ordinary far-infrared
absorption, or equivalently, the real part of the frequency-dependent
conductivity.  We consider two temperatures: $T \sim 2$ and $20$K.
At very low temperatures, the absorption spectra for
assumption (\ref{eq:ratioSimple}) resemble a sum of three single-gap 
spectra which become nonzero at the three different possible values
of $\Delta_i + \Delta_j$. If the matrix elements are diagonal,
the spectra would resemble the sum of {\em two} single-gap
spectra.  In either case, the lowest frequency where 
absorption occurs is determined by the smallest gap, which we 
denote $\Delta_{i_0}$, and occurs at frequency 
$\omega = 2\Delta_{i_0}/\hbar$.  Other absorption
edges, where $\alpha_S/\alpha_N$ has a slope discontinuity 
at $T = 0$, occur
at other values of $\Delta_i + \Delta_j$, but are not very
prominent, especially for the case II spectra. 
For the present model, the 
smaller gap $\Delta_{3D}\approx 2.5$meV, and the corresponding
absorption edge is at  $2\Delta_{3D}/k_BT_c\approx 1.4$, well below
the BCS value.

\begin{figure}
 \leftline{
 \epsfig{file=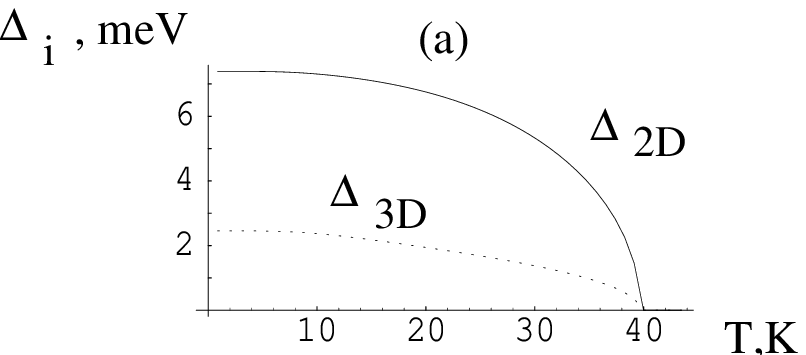, width=2.6in}
 }
 \vskip0.7pc
 \centerline{
 \epsfig{file=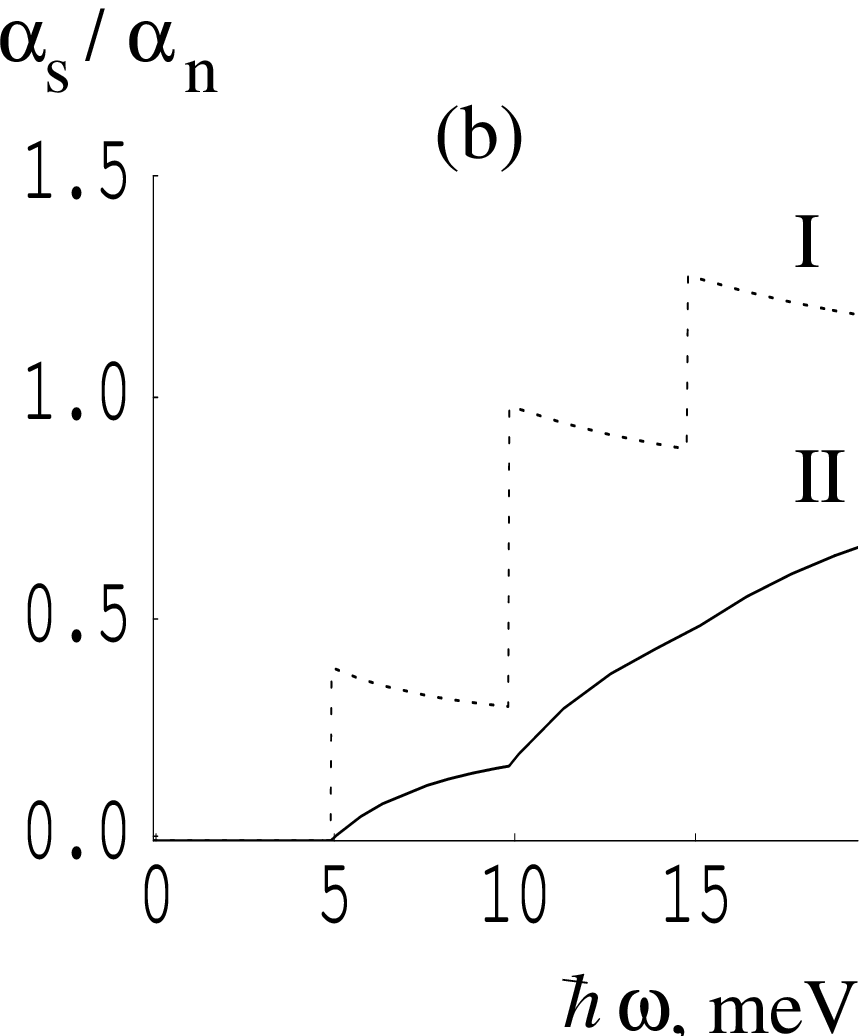, width=1.4in}
 \hskip1.5pc
 \epsfig{file=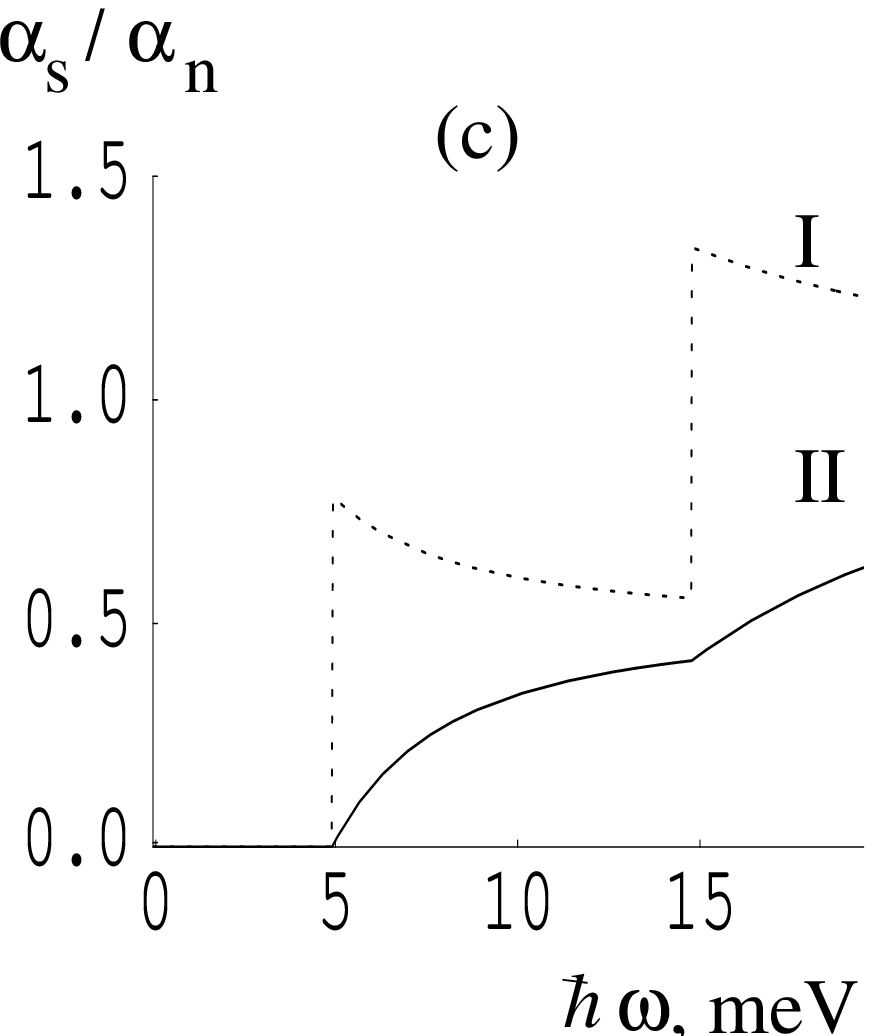, width=1.4in}
 }
 \vskip1pc
 \centerline{
 \epsfig{file=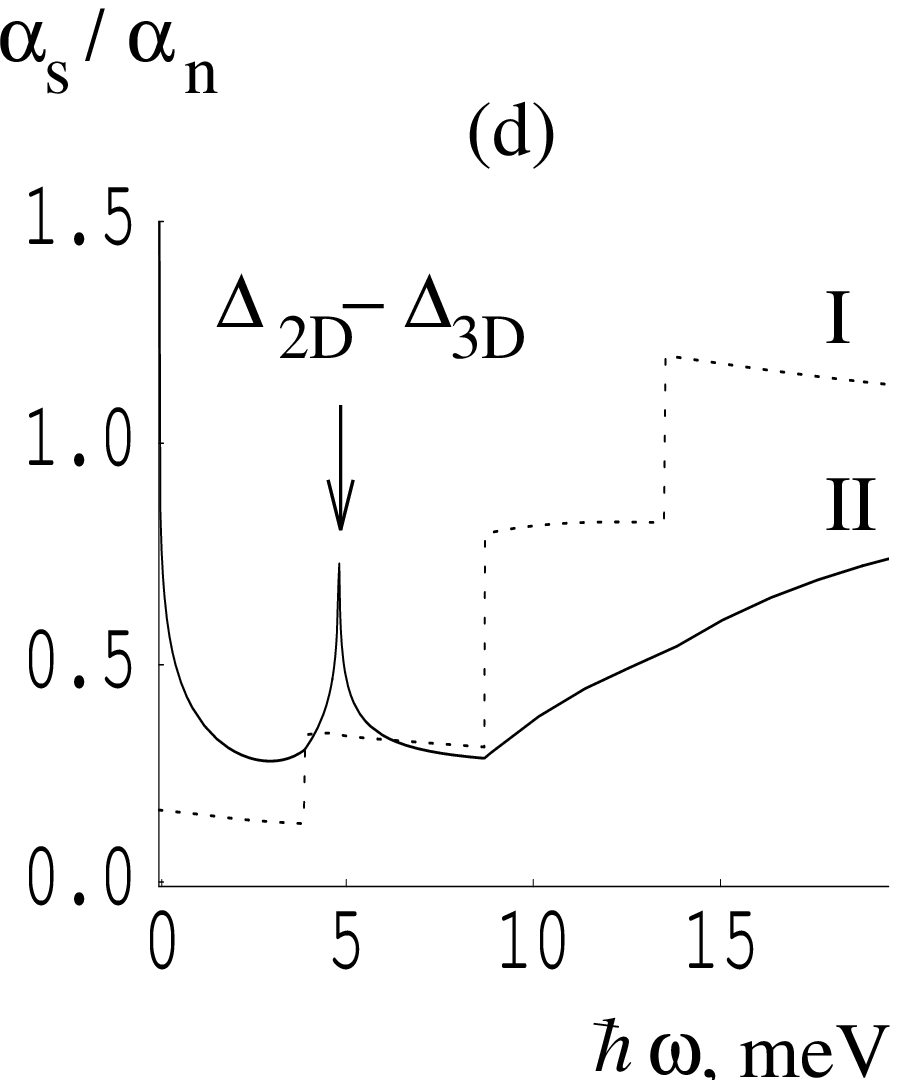, width=1.4in}
 \hskip1.5pc
 \epsfig{file=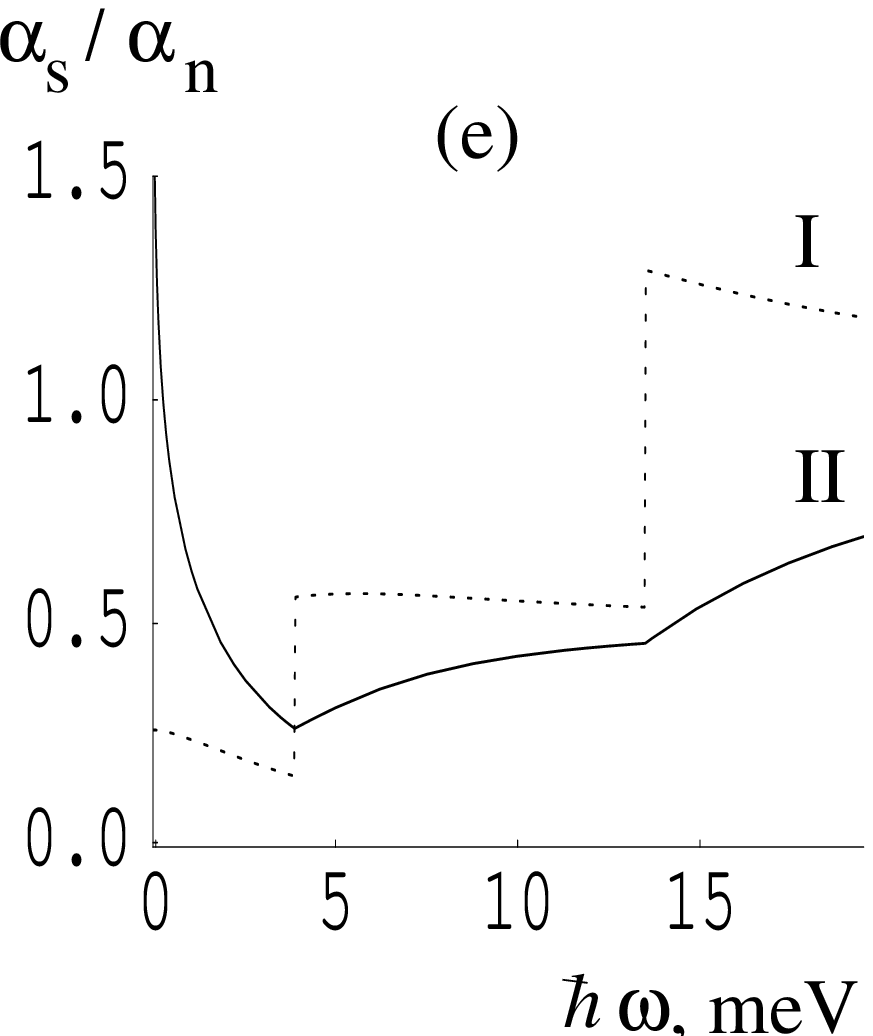, width=1.4in}
 }
 \vskip 1pc
\caption{
(a) The temperature dependence of the two gaps for the model 
discussed in the text; (b) and (c) The frequency dependence of the 
case I and case II absorption coefficients for the same model
(normalized to the normal state values) at $T = 0.05 T_c$ assuming
(b) eq.\ (\ref{eq:ratioSimple}) and (c) eq.\ (\ref{eq:ratioDiag}) 
for the transition matrix elements;
(d) and (e) same as (b) and (c) but with $T = 0.5 T_c$.}
\end{figure}

The case II absorption
coefficients at finite temperatures [Fig.\ 1(c) and 1(d)]  
show features not present in the lowest temperature plots, whether
the transition matrix elements are assumed to obey
eq.\ (\ref{eq:ratioSimple}) or eq.\ (\ref{eq:ratioDiag}).
First, there is a weak below-gap absorption for
all frequencies $\hbar \omega<2\Delta_{i_0}$.
This absorption is also present in the single-gap case.
But in the multiple gap superconductor, this background absorption, 
together with the peak feature discussed below, may mask the 
case II absorption edge at $2\Delta_{i_0}$.  (This edge is
still visible for case I spectra, even at finite T.)  
Instead, for case II spectra, 
one of the secondary edges at $\Delta_i + \Delta_j >2\Delta_{i_0}$ 
may become more prominent than the minimum absorption edge.
In addition, if the {\em nondiagonal} matrix elements are
substantial as in eq.\ (\ref{eq:ratioSimple}), 
there is an extra peak at the frequency of the {\em gap difference}, 
$\hbar \omega_{ij}= |\Delta_i-\Delta_j|$.   
At this frequency, the
integrand in eq.\ (\ref{eq:Iij}) has two multiplicative
singularities [square roots in the denominator of (\ref{eq:Iij})] 
As a result, the $\alpha_S$ calculated from eqs.\ (3) and (4)
is proportional to $-\ln |\omega - \omega_{ij}|$), with
a coefficient which is proportional to the number of 
thermally excited quasiparticles. 
However, if the matrix elements satisfy the diagonality assumption
(\ref{eq:ratioDiag}), this extra peak is absent.  
The origin of this peak(s) is the same as 
that of the better-known peak at $\omega=0$ in the single-gap case.
This latter peak is responsible
for the rise of the nuclear relaxation rate
$1/T_1$ to a value exceeding the normal-state value as the 
superconductor is cooled through $T_c$.  As in the $1/T_1$ case,
the actual height and width of the peak
at $|\Delta_i-\Delta_j|$ can be determined 
only when the 
${\bf k}$-dependence of the superconducting gaps, 
$\Delta_i({\bf k})$,
is included.  

We now move on to discuss the recent measurements 
of the infrared conductivity in \mb films\cite{mbOrenstein}.
In Fig.\ 2 (adapted from Fig.\ 2 of Ref.\cite{mbOrenstein}),
we show the measured real part of the optical conductivity,
$\sigma_1(\omega)$, normalized to its value in the normal state
at 40K, $\sigma_{1N}(\omega)$. 
These data show several unusual features.  
First, there is an apparent 
absorption edge at
5meV (well below both the weak- and strong-coupling single-gap 
BCS values), and a large background absorption below this edge.
Second, the experimental data show a weaker and more slowly rising 
$\sigma_{1S}/\sigma_{1N}$ ratio 
above the apparent absorption edge than a single-gap Mattis-Bardeen 
calculation
(shown for $2\Delta=5$ meV and $T=6$ K as a dashed line).
Finally, these spectra appear to have a characteristic structure
between 3 and 4 meV, present only in the superconducting state and 
reminiscent
of the peak structure discussed above. 
However, this experimentally observed structure could lie within
the experimental uncertainty, which is largest at 
low frequencies, where it is comparable to the scatter in the data
points\cite{OrensteinPrivate}.   Indeed, earlier data
by Pronin \etal\cite{mbPronin} is inconclusive regarding the
presence of such 3-4 meV structure.
More accurate measurements are 
needed to determine whether the peak structure is actually
present in the \mb spectra.

Clearly, the simple model of
in Fig.\ 1 cannot be directly applied 
to the measurements shown in Fig.\ 2.  First, the 
the $\omega_D$ parameter of the BCS model (\ref{eq:BCS}) was 
chosen to give $T_c\sim 40$ K, whereas the sample of
Fig.\ 2 has $T_c \sim 30.5$ K.   
At minimum, we should
multiply all the energy and temperature parameters
used to obtain Fig.\ 1 by a factor of $\sim 3/4$,  
to account for this lower $T_c$.
Also, the factors $M_{ij}N_i(0)N_j(0)$ need not be independent
of $i$ and $j$, as was assumed for Fig.\ 1 (b) and (c), and
the factors $M_{ii}N_i(0)^2$ need not be independent of
$i$ as assumed in Fig.\ 1(d); these factors could
also depend on the polarization of the 
radiation. 
Finally, the actual distribution of gap values should be included.
In the clean limit this distribution [$\Delta_i({\bf k})$] 
is predicted to be 
broad and to have several peaks in both ``2D'' and ``3D'' 
regions\cite{mb111823}.  But
impurity scattering should lead to ``averaging'' of the gap values
in the experimental sample ($\ell\sim100$ \AA).  The
resulting gap distribution depends strongly on
the relative magnitudes of various scattering 
rates, and in \mb these rates are thought to be 
small for scattering {\em between} 2D and 3D
states\cite{MazinCleanDirty,GolubovMazin}, 
thus justifying the use of a ``two-gap'' model, as mentioned
above. 
But for relatively small impurity concentrations, the
gap values should still have some dispersion
around the average $\Delta_{2D}$ and $\Delta_{3D}$.
The clean-limit results of Ref.\ \cite{mb111823} suggest
that two slightly different gaps, $\Delta_{2D,1}$ and
$\Delta_{2D,2}$, might form on the two 2D Fermi
surfaces if the {\em intraband} scattering is stronger than
scattering between two different 2D bands, for relatively clean
samples.

To illustrate these effects, 
we have calculated $\alpha_S/\alpha_N$ for case
II coherence factors using eqs.\ (4) and (5), with
parameters {\em arbitrarily} chosen so as to give a reasonable fit
between the calculated results and the measured $\alpha_S/\alpha_N$.
To illustrate possible effects of nontrivial gap value distribution,
we used either two gaps ($\Delta_{2D}$ and $\Delta_{3D}$), or
three gaps ($\Delta_{2D,1}$, $\Delta_{2D,2}$ and $\Delta_{3D}$).
In both cases, we have assumed that the off-diagonal matrix
elements are substantial\cite{note:3gap}.
The calculated curves shown in the insets of Fig.\ 2 (solid line)
have most of the features of experiment:
an apparent absorption
edge at a frequency well below the single-gap BCS value; 
a large background absorption below
this edge\cite{note:edge} at finite $T$;
and a more slowly rising $\alpha_S(\omega)/\alpha_N(\omega)$ with
increasing frequency above the
absorption edge than is predicted by the single-gap BCS model
(inset of Fig.\ 2, dashed line).  
However, we emphasize that the extra sharp peaks in the insets
are present only if the off-diagonal matrix elements $M_{ij}$
are substantial, which may well not be the case in MgB$_2$.

In conclusion, we have presented a simple calculation of far-infrared
absorption in a model for MgB$_2$, using a generalization of the
Mattis-Bardeen formula to several gaps.  The resulting absorption
ratio $\alpha_S(\omega)/\alpha_N(\omega)$ shows a 
qualitative resemblance to the measured results.  
Specifically,
the onset of absorption at a frequency well below the
onset predicted by isotropic BCS theory, large background absorption
below the apparent absorption edge, and a slower rise of 
$\alpha_S(\omega)/\alpha_N(\omega)$ than in the single-gap case
are all easily reproduced by the multiple-gap model.
Thus, the recent optical conductivity
measurements\cite{mbOrenstein} appear to be consistent with
the hypothesis of multiple superconducting gaps in \mb.  If
certain interband transition matrix elements are sufficiently
large, the multiband model would lead to a peak
at a frequency corresponding to the gap difference 
$\Delta_{2D} - \Delta_{3D}$, which might
be observable at finite temperatures in moderately dirty 
samples.

This work has been supported by NSF through Grant DMR 01-04987 and
the U. S.-Israel Binational Science Foundation, 
and also benefited from the computational facilities of the
Ohio Supercomputer Center.  
We thank T.\ R.\ Lemberger for valuable
conversations.

\begin{figure}
\centerline{
\epsfig{file=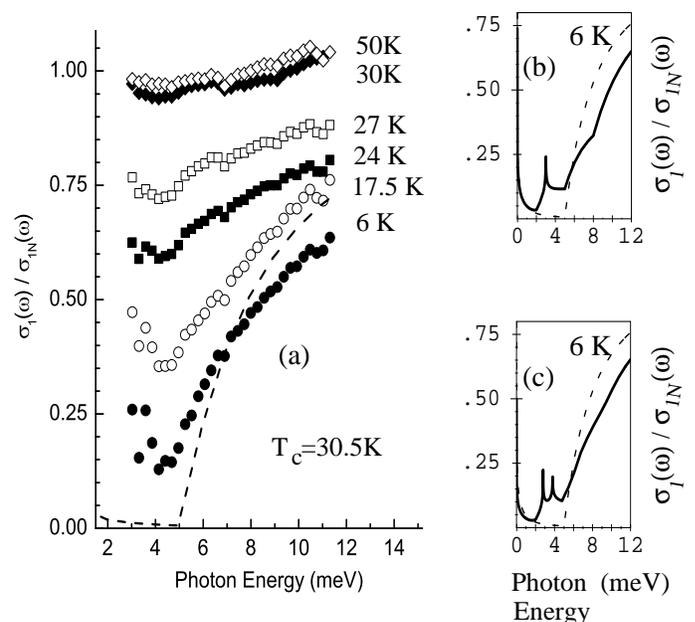, width=3.5in}
}
\caption{(a) Symbols: experimental $\sigma_1(\omega)$, 
scaled to the 40K value in the normal state,
as measured on a 100nm-thick film of \mb ($T_c\approx 30.5$ K) [adapted
from Ref.\ 13].  The dashed line shows the fit of the 6K data
to the single-gap Mattis-Bardeen expression (with $2\Delta=5$ meV),
as given in [13].  (b),(c) Calculated 
$\sigma_1(\omega)/\sigma_N(\omega)$ for (b) a two-gap or (c) a three-gap 
model, assuming nonzero off-diagonal transition matrix
elements as discussed in the text (solid lines), and for a BCS
single-gap model (dashed lines) using $2\Delta = 5$ meV.
}
\end{figure}


\begin{thebibliography}{99}

\bibitem{Liu}A. Y. Liu, I. I. Mazin, and J. Kortus
Phys. Rev. Lett. {\bf 87}, 087005 (2001).

\bibitem{mb111823}H.J. Choi, D. Roundy, H. Sun, M. L. Cohen, S. G. Louie,
cond-mat/0111182; cond-mat/0111183.

\bibitem{note:2D} Both $\Delta_{2D}({\bf k})$ 
and $\Delta_{3D}({\bf k})$
are calculated to be essentially isotropic (i. e. only weakly
dependent on ${\bf k}$);
the subscripts here refer to the topologies 
of the corresponding Fermi surfaces.

\bibitem{mb07072}  F. Bouquet, R. A. Fisher, N. E. Phillips, D. G.
Hinks, and J. D. Jorgensen,
Phys. Rev. Lett. {\bf 87}, 047001 (2001);
R. A. Fisher, F. Bouquet, N. E. Phillips, D. G. Hinks, and J. D. Jorgensen,
in "Studies of High Temperature 
Superconductors," Vol.\ 38, Ed.\ A.\ V.\ Narlikar,
Nova Science, NY (2001);
F. Bouquet, Y. Wang, R. A. Fisher, D. G. Hinks, J. D. Jorgensen, A. Junod,
and N. E. Phillips, 
Europhysics Letters {\bf 56}, 856 (2001).

\bibitem{mb11262} A. A. Golubov, J. Kortus, O. V. Dolgov, cond-mat/0111262.

\bibitem{mb12144} H. Schmidt, J.F. Zasadzinski, K.E. Gray, 
D.G. Hinks, cond-mat/0112144.

\bibitem{mb12452} 
Yu. G. Naidyuk {\it et al}, cond-mat/0112452.

\bibitem{mb11115} A. Brinkman, A.A. Golubov, H. Rogalla, O. V. 
Dolgov, and J. Kortus, 
cond-mat/0111115.

\bibitem{mb0201200} A.I. D'yachenko, V.Yu. Tarenkov, A. V.
Abal'oshev, and S. J. Lewandowski,
cond-mat/0201200.

\bibitem{tsuda01} S. Tsuda, T. Yokoya, T.Kiss, Y. Takano, K. Togano,
H. Kito, H. Ihara, and S. Shin, Phys. Rev. Lett. 
{\bf 87}, 177006 (2001).

\bibitem{mbLemberger} M.-S. Kim, J.A. Skinta, T.R. Lemberger, W. N.
Kang, H.-J. Kim, E.-M. Choi, and S.-I. Lee,
cond-mat/0201550.

\bibitem{MattisBardeen} D.C. Mattis and J. Bardeen, 
Phys. Rev. {\bf 111}, 412 (1958).

\bibitem{MazinCleanDirty} I.I. Mazin, O.K. Andersen, O. Jepsen, O.V. Dolgov, 
J. Kortus, A.A. Golubov, A.B.Kuz'menko and D. van der Marel,
cond-mat/0204013.

\bibitem{mbOrenstein}R.A. Kaindl, M.A. Carnahan, J. Orenstein, and
D. S. Chemla,
Phys. Rev. Lett., {\bf 88}, 027003 (2002).

\bibitem{referee1a} S.\ B.\ Nam, Phys. Rev., {\bf 156}, 470 and 487 
(1967); P.\ B.\  Allen, Phys.\ Rev. {\bf B 3}, 305 (1971).

\bibitem{referee1b} O. Klein, E.\ J. Nicol, K. Holczer and 
G.\ Gr\"uner, Phys. Rev. {\bf B 50}, 6307 (1994).

\bibitem{note:model} From the parameters of Ref. \cite{Liu},
the products $\Lambda_{ij}\equiv U_{ij}N_j(0)$

take the values $\Lambda_{11}=0.96$, $\Lambda_{12}=0.16$,
$\Lambda_{21}=0.22$, $\Lambda_{22}=0.28$, with ``1'' referring to the 2D
and ``2'' to the 3D bands. The values of $T$, $\Delta_i$ and 
$\omega$ are
then all scaled by
a single BCS cutoff parameter $\omega_D$. 
$T_c \approx 40$K corresponds to $\omega_D \approx 7.5$ meV.

\bibitem{Tinkham} M. Tinkham, Introduction to Superconductivity, 2nd. ed.,
McGraw-Hill, 1996; pp. 79-89.

\bibitem{Yildirim} T. Yildirim \etal,
Phys. Rev. Lett. {\bf 87}, 037001 (2001).

\bibitem{note:cancel} To calculate corrections resulting from 
any noncancellation of linear terms in (\ref{eq:factor}) would
require a treatment more elaborate than the BCS 
picture discussed here.


\bibitem{Masuda62} Y. Masuda, Phys. Rev. {\bf 126}, 1271 (1962).

\bibitem{OrensteinPrivate} R.A. Kaindl, private communication.

\bibitem{mbPronin} A.V. Pronin, A. Pimenov, A. Loidl, S.I. Krasnosvobodtsev,
Phys. Rev. Lett {\bf 87}, 097003 (2001).

\bibitem{GolubovMazin} A.A. Golubov and I.I. Mazin, Phys.Rev. {\bf B 55},
15146 (1997).

\bibitem{note:3gap} For the calculations in the insets of the Figure 2 
we used the following parameters:
For inset (b), $\Delta_{2D}=4$ eV,
$\Delta_{3D}=1$ eV;
for the matrix elements 
(in units such that $|M_{2D,2D}|^2=1$): $|M_{2D,3D}|^2=.25$,
$|M_{3D,3D}|^2=0.125$.
For inset (c),
$\Delta_A=3.8$, $\Delta_B=4.8$, $\Delta_C=1$.
(We use the compressed notation
$A\equiv 2D,1$; $B\equiv 2D,2$; $C\equiv 3D$.) 
For the matrix elements
(in units such that $|M_{AA}|^2=1$): $|M_{AB}|^2=2$,
$|M_{AC}|^2=1$, $|M_{BB}|^2=10$, $|M_{BC}|^2=2$, $|M_{CC}|^2=0.3$.
The values of $N_A(0)$, $N_B(0)$
and $N_C(0)$ were taken from Ref.[1]
(1.38, 0.66 and 2.78 states/[Ry-spin-primitive cell]).

\bibitem{note:edge} The apparent 
absorption edge at $\sim 5$ meV in our fitted model is actually due 
to transitions between the 2D and 3D bands that begin
at $\Delta_{2D}+\Delta_{3D}$ (or $\Delta_{2D,1}+\Delta_{3D}$).
The true $T=0$ absorption edge is at $2 \Delta_{3D}=2$ meV 
and is responsible for
the large background below 5 meV, again consistent with experiment.


\end{thebibliography}
\end{document}